      [1999/12/01 v1.4c Il Nuovo Cimento]
\documentclass{cimento}

\usepackage{graphicx}  
\title{Search for Dark Matter in the sky with the Fermi Large Area Telescope}
\author{A.~Morselli \from{ins:x}\ETC,
E.~Nuss \from{ins:evil},
G.~Zaharijas \from{ins:ee}
}
\instlist{\inst{ins:x} INFN Roma Tor Vergata
  \inst{ins:evil} Laboratoire Univers et Particules de Montpellier, Universit\'e Montpellier 2, Montpellier
   \inst{ins:ee} The Abdus Salam International Centre for Theoretical Physics and INFN, Trieste}
\PACSes{PACS 96.50.sb, 95.35.4-d, 95.85.Ry, 98.70.Sa}

\begin{document}

\maketitle

\begin{abstract}
Can we learn about New Physics with astronomical and astro-particle data?
 Since its launch in  2008, the Large Area Telescope, onboard of the  Fermi Gamma-ray Space Telescope, has detected the largest amount of gamma rays in the 20 MeV - 300 GeV energy range and electrons + positrons in the 7 GeV- 1 TeV range. These impressive statistics allow one to perform a very sensitive indirect experimental search for dark  matter. We will present the latest results on these searches and the comparison with LHC searches.\end{abstract}

\section{Introduction}

\begin{figure}[htb]
\centerline{%
\includegraphics[width=12.5cm]{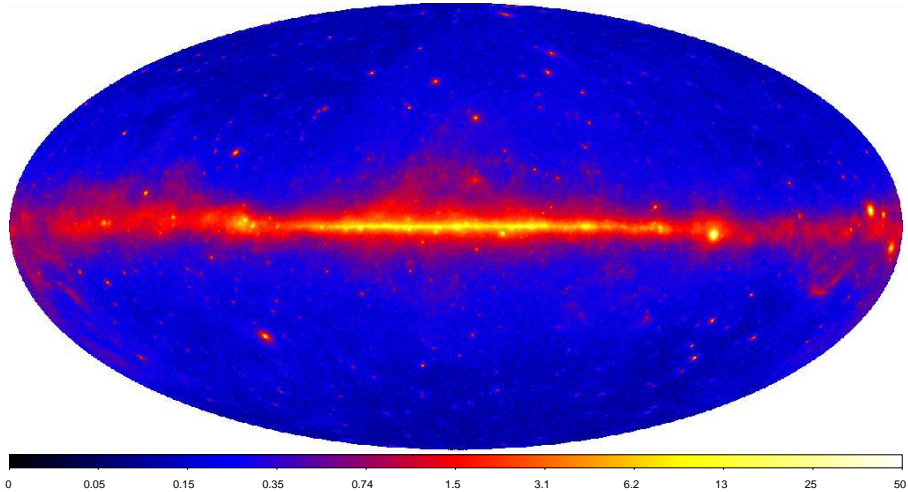}}
\caption{Sky map of the energy flux derived from 24 months of observation. The image shows $\gamma$-ray energy flux for energies between 100~MeV and 10~GeV, in units of 10$^{-7}$ erg cm$^{-2}$ s$^{-1}$ sr$^{-1}$.}
\label{eflux_map}
\end{figure}

The Fermi Observatory carries two instruments on-board: the Gamma-ray Burst Monitor (GBM) \cite{GMB}  and the Large Area Telescope (LAT) \cite{Fermi_rev}. 
The LAT is a pair conversion telescope for photons above 20 MeV up to a few hundreds of GeV. The field of view is $\sim $2.4 sr and LAT observes the entire sky every $\sim $ 3 hours (2 orbits). These features make the LAT a great instrument for dark matter (DM) searches.
The operation of the instrument through the first three years of the mission was smooth at a level which is probably beyond the more optimistic pre- launch expectations. The LAT has been collecting science data for more
than 99\% of the time spent outside the South Atlantic Anomaly (SAA). The remaining tiny fractional down-time accounts for both hardware issues and detector calibrations \cite{calib}, \cite{calib2}.

More than 650 million gamma-ray candidates (i.e. events passing the background rejection selection) were made public and distributed to the Community through the Fermi Science Support Center (FSSC) \footnote{The FSSC is available at {\sl http://fermi.gsfc.nasa.gov/ssc}}.

Over the first three years of mission the LAT collaboration has put a considerable effort toward a better understanding of the instrument and of the environment in which it operates. In addition to that a continuous effort was made to in order to make the advances public as soon as possible. In August 2011 the first new event classification (Pass 7) since launch was released, along with the corresponding Instrument Response Functions ( and a release of a new event class 'Pass 7 reprocessed' is planned for the near future). Compared with the pre-launch (Pass 6 ) classification, it features a greater and more uniform exposure, with a significance enhancement in acceptance below 100 MeV.

\section{The Second Fermi-LAT catalog}

The high-energy gamma-ray sky is dominated by diffuse emission: more than 70\% of the photons detected by the LAT are produced in the interstellar space of our Galaxy by interactions of high-energy cosmic rays with matter and low-energy radiation fields. An additional diffuse component with an almost-isotropic distribution (and therefore thought to be extragalactic in origin) accounts for another significant fraction of the LAT photon sample. The rest consists of various different types of point-like or extended sources: Active Galactic Nuclei (AGN) and normal galaxies, pulsars and their relativistic wind nebulae, globular clusters, binary systems, shock-waves remaining from supernova explosions and nearby solar-system bodies like the Sun and the Moon.

The Second Fermi-LAT catalog (2FGL) \cite{2FGL} is the deepest catalog ever produced in the energy band between 100 MeV and 100 GeV. Compared to the First Fermi-LAT (1FGL) \cite{1FGL} , it features several significant improvements: it is based on data from 24 (vs. 11) months of observation and makes use of the new Pass 7 event selection. The energy flux map is shown in  figure~\ref{eflux_map} .   It is interesting to note that 127 sources are firmly identified, based either on periodic variability (e.g. pulsars) or on spatial morphology or on correlated variability. In addition to that 1170 are reliably associated with sources known at other wavelengths, while 576 (i.e. 31\% of the total number of entries in the catalog) are still unassociated. 
 In addition, the first catalog of  high energy sources   \cite{Paneque:2013aba}  as well as the first SNR catalog are in preparation \cite{dePalma:2013pia}.

\section{Indirect Dark Matter searches}

One of the major open issues in our understanding of the Universe is the existence of an extremely-weakly interacting form of matter, the Dark Matter (DM), supported by a wide range of observations including large scale structures, the cosmic microwave background and the isotopic abundances resulting from the primordial nucleosynthesis. Complementary to direct searches being carried out in underground facilities and at accelerators, the indirect search for DM is one of the main items in the broad Fermi Science menu.
The word indirect denotes here the search for signatures of Weakly Interactive Massive Particle (WIMP) annihilation or decay processes through the final products (gamma-rays, electrons and positrons, antiprotons) of such processes. Among many other ground-based and space-borne instruments, the LAT plays a prominent role in this search through a variety of distinct search targets: gamma-ray lines, Galactic and isotropic diffuse gamma-ray emission, dwarf satellites, CR electrons and positrons.

\subsection{Galactic center}
The Galactic center (GC) is expected to be the strongest source of $\gamma$-rays from DM 
annihilation, due to its coincidence with the cusped part of the DM halo density profile \cite{dark1},  \cite{dark}, \cite{dark2}.
A preliminary analysis of the data, taken during the first 11 months of the Fermi satellite operations is presented in 
 \cite{F_sym}, \cite{GC_cim}.

The diffuse gamma-ray backgrounds and discrete sources, as we know them today, can account for the large majority of the 
detected gamma-ray emission from the Galactic Center. Nevertheless a residual  emission is left, not accounted for by the above models \cite{F_sym}, \cite{GC_cim}. Improved modeling of the Galactic diffuse model as well as the potential contribution from other astrophysical sources (for instance unresolved point sources) could provide a better description of the data. Analyses are underway to investigate these possibilities.

\subsection{ Galactic halo}

\begin{figure}[t]
\begin{center}$
\begin{array}{cc}
\includegraphics[width=0.5\columnwidth]{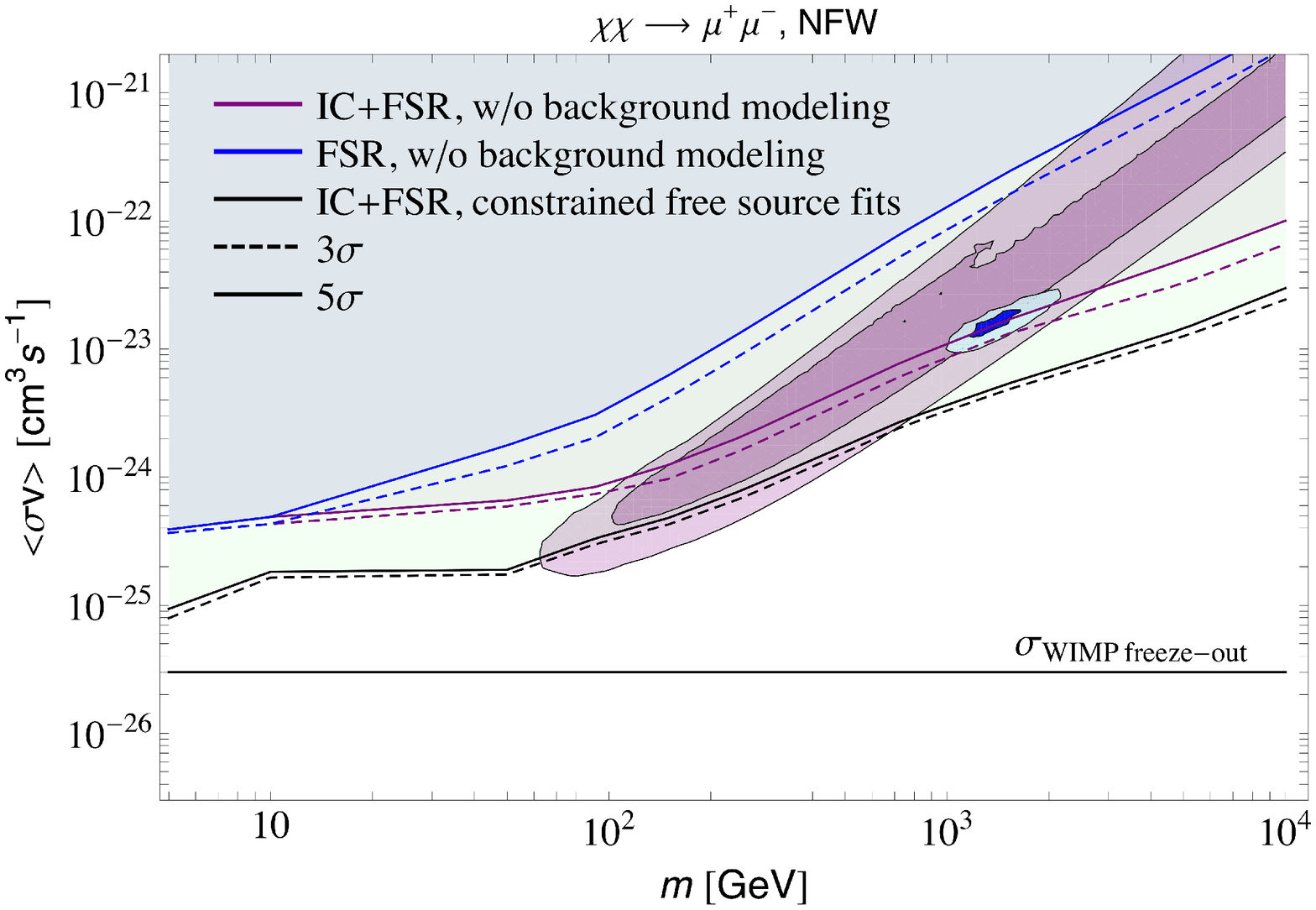}&
\includegraphics[width=0.5\columnwidth]{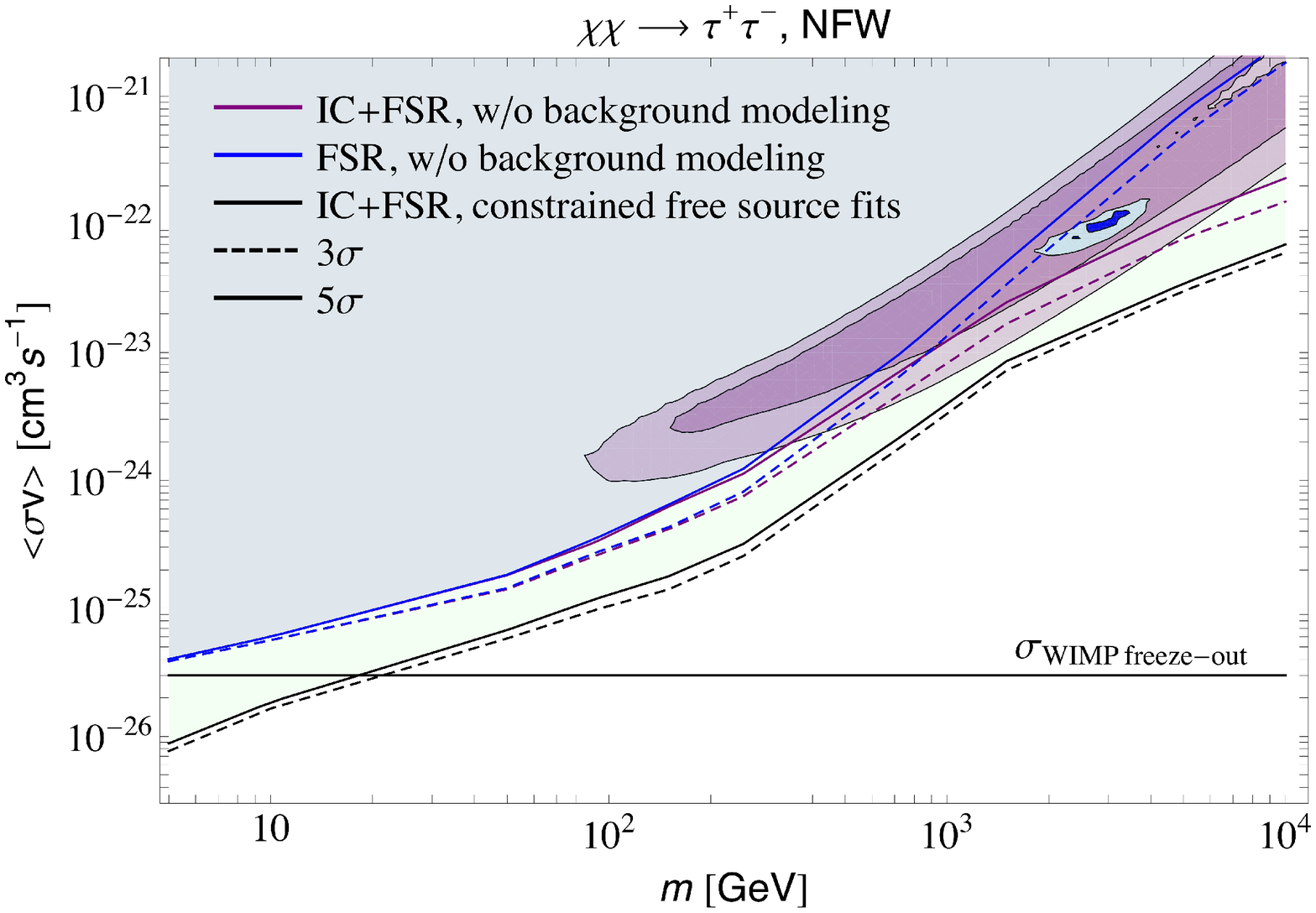} \\

\end{array}$
\end{center}
\vspace{-0.5cm}
\caption{Derived 95\% C.L. upper limits on WIMP annihilation cross sections in the Milky Way halo, for the muon ({\it left}) and tau ({\it right}) annihilation channels.}
\label{halo_mu}
\end{figure}

{ In order to minimize uncertainties connected with the region of the Galactic Center, analysis \cite{Halo} considered a region of interest consisting of two off-plane rectangles ($5^0~\leq |b|\leq ~15^0$ and $|l|\leq~80^0$) and searched for continuum emission from dark matter annihilation or decay in the smooth Galactic dark matter halo. They considered two approaches: a more conservative one in which limits were set on DM models assuming that all gamma ray emission in that region might come from dark matter (i.e. no astrophysical signal is modeled and subtracted). In a second approach, dark matter source and astrophysical emission was fit simultaneously to the data, marginalizing over several relevant parameters of the astrophysical emission. As no robust signal of DM emission is found, DM limits are set.

These limits are particularly strong on leptonic DM channels, which are hard to constrain in most other probes (notably in the analysis of the dwarf Galaxies, described below). This analysis strongly challenges DM interpretation \cite{Fermi_el_interpretation} of the positron rise, observed by PAMELA \cite{pamela_el_2011} and Fermi LAT \cite{Fermi_el,Fermi_el2} (see figure \ref{halo_mu}).}

\begin{figure}[t]
\begin{center}$
\begin{array}{cc}
\includegraphics[width=0.5\columnwidth]{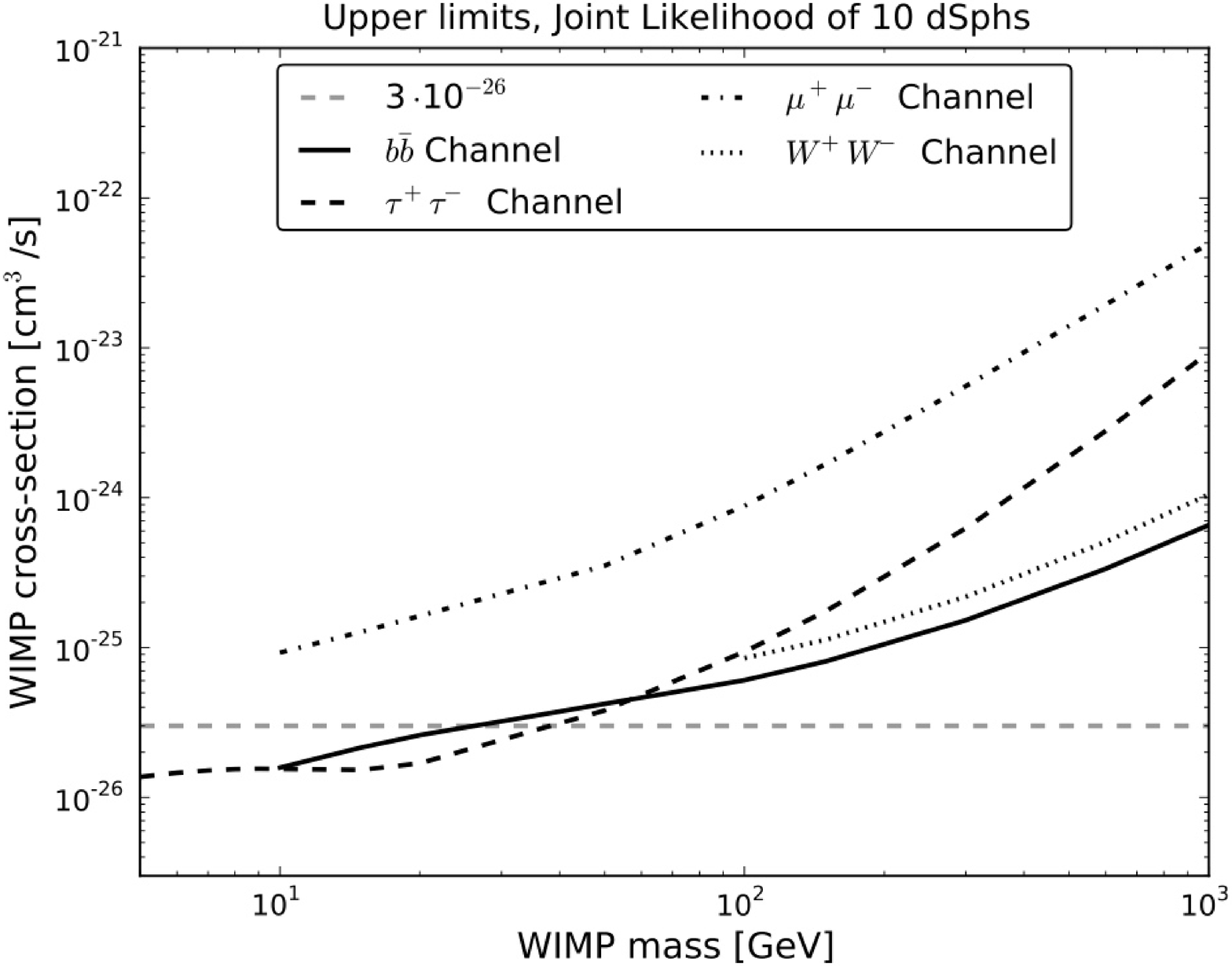}&
\includegraphics[width=0.5\columnwidth]{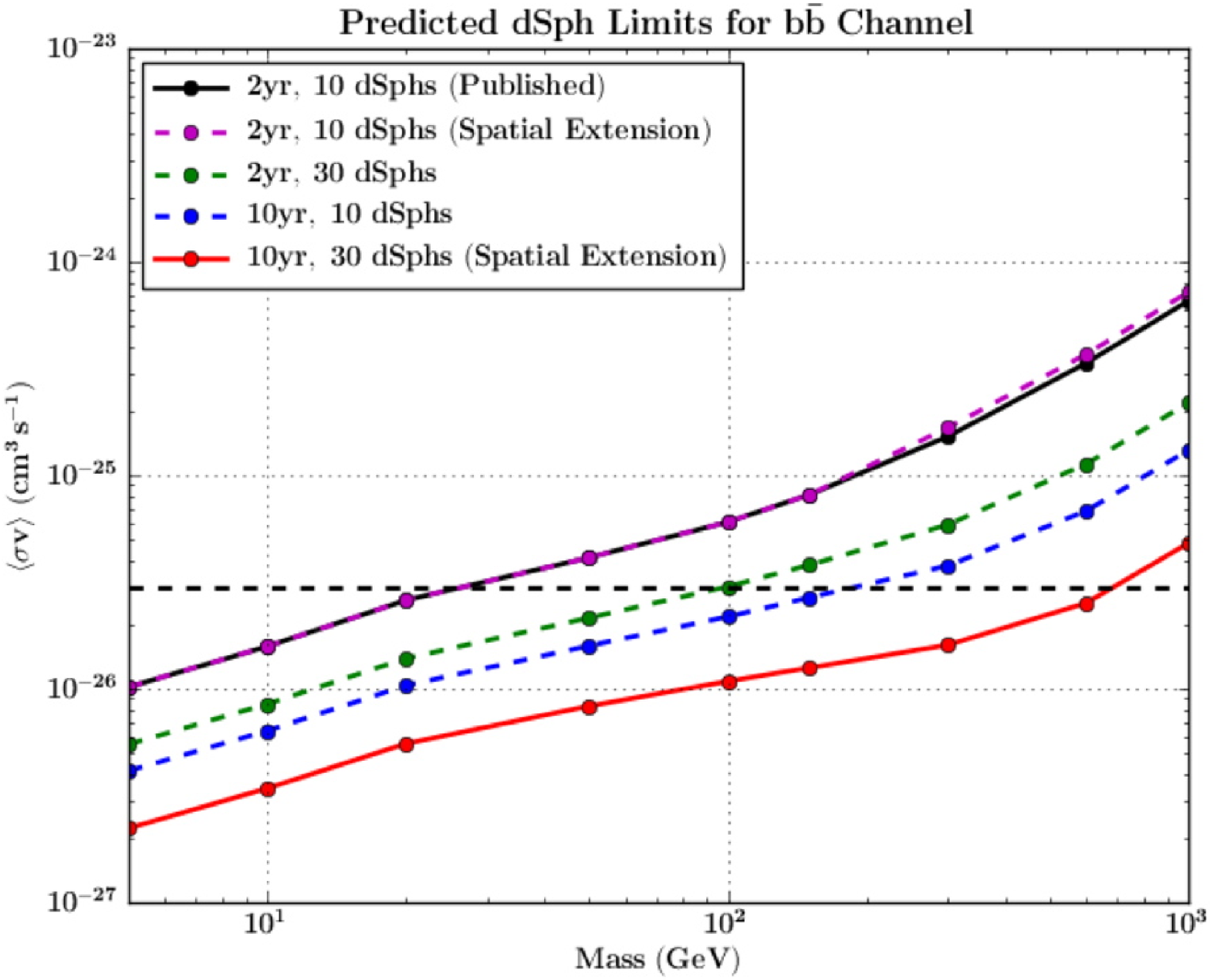} \\
\end{array}$
\end{center}
\vspace{-0.5cm}
\caption{{\it Left:} Derived 95\% C.L. upper limits on WIMP annihilation cross sections for different channels. {\it Right:} Predicted 95\% C.L. upper limits on WIMP annihilation cross sections in 10 years for $ b bar $ channel.}
\label{Dwarf_fig}
\end{figure}

\subsection{ Dwarf galaxies}

Dwarf satellites of the Milky Way are among the cleanest targets for indirect dark matter searches in gamma-rays. They are systems with a very large mass/luminosity ratio (i.e. systems which are largely DM dominated). The LAT detected no significant emission from any of such systems and the upper limits on the $\gamma$-ray flux allowed us to put very stringent constraints on the parameter space of well motivated WIMP models \cite{Dwarf}.

A combined likelihood analysis of the 10 most promising dwarf galaxies, based on 24 months of data and pushing the limits below the thermal WIMP cross section for low DM masses (below a few tens of GeV), has been recently performed  \cite{Dwarf2}. The main advantages of the combined likelihood are that the analysis can be individually optimized and
that combined limits are more robust under individual background fluctuations and under individual
astrophysical modelling uncertainties than individual limits. The derived 95\% C.L. upper limits on WIMP annihilation cross sections for different channels are shown in figure \ref{Dwarf_fig} (left). The  most  generic cross section ($\sim  3 \cdot 10^{-26} cm^3 s^{-1} $ for a purely s-wave cross section) is plotted as a reference.  These results are obtained for NFW profiles \cite{nfw} but for  cored dark matter profile the J-factors for most of the dSphs would either increase or not change much so these results includes J-factor uncertainties   \cite{Dwarf2}. 

With the present data   we are able to rule out large parts of the parameter space where the thermal relic density is below the observed cosmological dark matter density and WIMPs  are dominantly produced non-thermally, e.g. in models where supersymmetry
breaking occurs via anomaly mediation
 for the MSSM model, updated from \cite{Dwarf}).

Future improvements  (apart from increased amount of data) will include an improved event selection with a larger effective area and photon energy range, and the inclusion of more satellite galaxies.
In figure \ref{Dwarf_fig} (right) 
are shown  the predicted upper limits in the hypothesis of  10 years of data instead of 2; 30 dSphs instead of ten (supposing that the new optical surveys will find new dSph);  spatial extension analysis (source extension increases the signal region at high energy$ E \ge 10 ~GeV, M  \ge  200 ~GeV$ ).

Other   complementary  limits  were obtained with  the search of possible  anisotropies generated by the DM halo substructures  \cite{ani2},  the search for Dark Matter Satellites  \cite{DarSat} 
and a  search for high-energy cosmic-ray electrons from the Sun \cite{DarkSun}.

\subsection{Gamma-ray lines}

A line at the WIMP mass, due to the 2$\gamma$ production channel, could be observed as a feature in the astrophysical source spectrum \cite{dark2}. Such an  observation would be  a ``smoking gun'' for WIMP DM as it is difficult to explain by a process other than WIMP annihilation or decay and the presence of a feature due to annihilation into $\gamma Z$ in addition would be even more convincing.  No significant evidence of gamma-ray line(s) has been found in the first two years of data  from 7 to 200 GeV \cite{lines2} (see also \cite{lines}).

Recently, the claim of an indication of line emission in Fermi-LAT data  \cite{lineW,lineF}  has drawn considerable attention.
Using an analysis technique similar to \cite{lines}, but doubling the amount of data as well as optimizing the region of interest for signal over square-root of background, \cite{lineW} found a (trial corrected) 3.2~$\sigma$ significant excess at a mass of  $\sim 130$~GeV  that, if interpreted as a signal would amount to a cross-section of about $<\sigma v > \sim 10^{-27} cm^3s^{-1}$. 

\begin{figure}
\centerline{
\includegraphics[height=8.5cm,width=13.5cm]{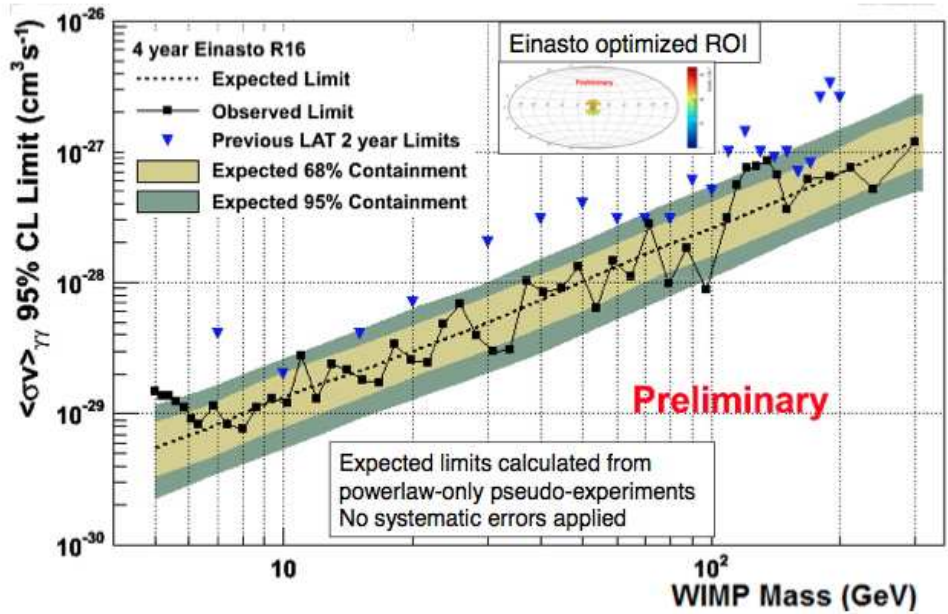}}
\caption{Dark matter annihilation 95\% CL cross section upper limits into $\gamma\gamma$  for the Einasto  profile for a circular region  of interest (ROI) with  a radius R$_{GC} = 16^\circ$ centered on the GC with $|b|<5^\circ$  and $|l|>6^\circ$ masked.
}
\label{fig:lines2}
\end{figure}

The signal is  found to be concentrated on the Galactic Centre with a spatial distribution consistent with an Einasto profile  \cite{Einasto}. This is marginally compatible with the upper limit presented in  \cite{lines2}. 
 In the analysis of the 4 year data the Fermi LAT team has improved over the two year paper in three important aspects: i) the search was performed in five regions of interest optimized for DM search under five different assumptions on the morphology of the DM signal, ii) new improved data set (pass 7 reprocessed) was used, as it corrects for loss in calorimeter light yield due to radiation damage during the four years of the Fermi mission and iii) point spread function (PDF) was improved by adding a 2nd dimension to the previously used triple Gaussian PDF model, leading to a so called '2D' PDF (such procedure is shown to increase the sensitivity to a line detection by $15\%$).. In that analysis \cite{line} no globally significant lines have been fond and new limits to this DM annihilation channel were set (see figure \ref{fig:lines2}). In a close inspection of the 130 GeV feature it was found that indeed there exist a 135 GeV signal at $4.01 \sigma$ local significance, when a '1D' PSF and old data sets were used (consistently with what \cite{lineW,lineF} have found). However, the significance drops to $3.35 \sigma$ (local, or $\leq2\sigma$ global significance once trials factors are taken into account). In addition, a weaker signal is found at the same energy in the control sample (in the Earth limb), which might point to a systematics effectpresent in this data set. In order to examine this possibility weekly observations of the Limb are scheduled, and a better understanding of a nature of the excess in the control sample should be available soon.

A new version of the event-level reconstruction and analysis framework (called Pass 8~) is foreseen  soon from the Fermi LAT collaboration. With this new analysis software we should increase the efficiency of the instrument at high energy and have a data set based on independent event analysis thus gaining a better control of the systematic effects.

\subsection{ The Cosmic Ray Electron spectrum}

The experimental information available on the Cosmic Ray Electron (CRE) spectrum has been dramatically expanded with a high precision measurement of the electron spectrum from 7 GeV to 1 TeV by the Fermi LAT \cite{Fermi_el}, \cite{Fermi_el2}. The spectrum  shows no  prominent spectral features and it is significantly harder than that inferred from several previous experiments. 

\begin{figure}
\centering
\includegraphics[height=9.5cm,width=12.5cm]{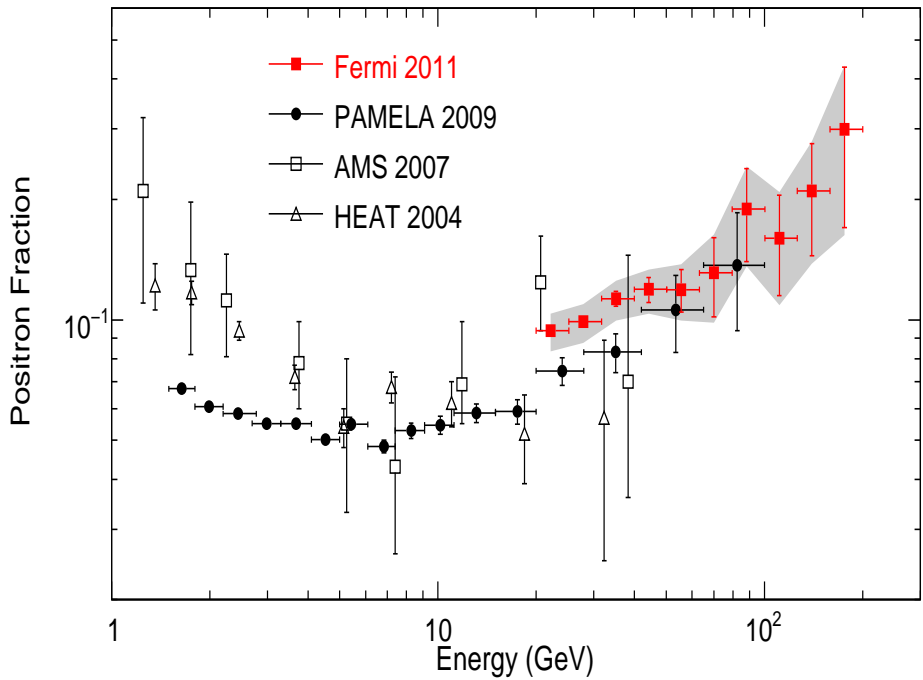}
\caption{Positron fraction measured by the Fermi LAT and by other experiments.
~\cite{HEAT2004, AMS2007, pamela_el_2011}. 
 The Fermi statistical uncertainty is shown with error bars and the total (statistical plus systematic uncertainty) is shown as a shaded band}
\label{PosFrc}
\end{figure}

Recently the Fermi-LAT collaboration performed a direct measurement of the absolute $e^+$ and $e^-$ spectra, and of their fraction \cite{Fermi_elpos}. As the Fermi-LAT does not carry a magnet, analysis took advantage of the fact that due to its magnetic field, the Earth casts a shadow in electron or positron fluxes in precisely determined regions. As a result, this measurement confirmed a rise of the positron fraction observed by PAMELA, between 20 and ~100 GeV  and determine for the first time that it continues to rise between 100 and 200 GeV (see figure \ref{PosFrc}).

These measurements show  that a new component of $e^+$ and $e^-$ are needed with a peak at  $\sim 1$ TeV.
The temptation to claim the discovery of dark matter  from detection of  electrons and positrons from annihilation of dark matter particles is strong
but  there are competing astrophysical sources, such as pulsars, that can give a strong 
flux of primary positrons and electrons
(see
 \cite{Fermi_el_interpretation} and references therein).
At energies between 100 GeV and 1 TeV the electron flux reaching the Earth may be the sum of an almost homogeneous and isotropic component produced by  Galactic supernova remnants and the local contribution of a few pulsars with the latter expected to contribute more and more significantly as the energy increases.
If a single nearby pulsar give the dominant contribution to the extra component a large anisotropy and a small bumpiness should be expected; if several pulsars contribute the opposite scenario is expected.   

So far no positive detection of CRE anisotropy was reported by the Fermi-LAT collaboration, but some stringent upper limits were published \cite{anis} and the pulsar scenario is still compatible with these upper limits. 

After the conference the AMS-02 collaboration presented the  result on the positron fraction \cite{AMS-02}
that confim the positron ratio rise observed by PAMELA and Fermi and extend it up to 350 GeV.

Forthcoming measurements from AMS-02 and CALET are expected to reduce drastically the uncertainties on the propagation parameters by providing more accurate measurements of the spectra of the nuclear components of CR. Fermi-LAT and those experiments are also expected to provide more accurate measurements of the CRE spectrum and anisotropy looking for features which may give a clue of the nature of the extra component.

\section{Conclusions}
Fermi turned four years in orbit on June, 2012, and it is definitely living up to its expectations in terms of scientific results delivered to the community. The mission is planned to continue at least four more years (likely more) with many remaining opportunities for discoveries.

\acknowledgments
The Fermi LAT Collaboration acknowledges support from a number of agencies and institutes for both development and the operation of the LAT as well as scientific data analysis. These include NASA and DOE in the United States, CEA/Irfu and IN2P3/CNRS in France, ASI and INFN in Italy, MEXT, KEK, and JAXA in Japan, and the K. A. Wallenberg Foun- dation, the Swedish Research Council and the National Space Board in Sweden. Additional support from INAF in Italy and CNES in France for science analysis during the operations phase is also gratefully acknowledged.


\begin{thebibliography}{0}
\bibitem{GMB} C.Meegan {\it et al.}, ApJ {\bf  702} (2009)  791 

\bibitem{Fermi_rev} W.B.Atwood {\it et al.} [Fermi Coll.],   ApJ {\bf 697}  (2009) 1071-1102       [arXiv:0902.1089]

\bibitem{calib}   M.Ackermann {\it et al.} [Fermi Coll.],  Astroparticle Physics {\bf 35} (2012) 346Ð353  [arXiv:1108.0201]

\bibitem{calib2}   M.Ackermann {\it et al.} [Fermi Coll.],  ApJS   {\bf203} (2012) 4  [arXiv:1206.1896]


\bibitem{2FGL} A.Abdo {\it et al.} [Fermi Coll.],     ApJS  {\bf 199}  (2012) 31  [arXiv:1108.1435]
\bibitem{1FGL}  A.Abdo {\it et al.} [Fermi Coll.],  ApJS {\bf 188} (2010) 405  [arXiv:1002.2280]

\bibitem{Paneque:2013aba} 
  D.~Paneque, J.~Ballet, T.~Burnett, S.~Digel, P.~Fortin and J.~Knoedlseder,
  arXiv:1304.4153 [astro-ph.HE].

\bibitem{dePalma:2013pia} 
  F.~de Palma {\it et al.}  [ for the Fermi LAT Collaboration],
  arXiv:1304.1395 [astro-ph.HE].

\bibitem{dark1} A. Morselli {\it et al.},  Nucl.Phys. {\bf 113B} (2002)  213 
\bibitem{dark} A.Cesarini, F.Fucito, A.Lionetto, A.Morselli, P. Ullio, Astropart.\ Phys.\  {\bf 21} (2004) 267  [astro-ph/0305075]

\bibitem{dark2}  E. Baltz {\it et al.}, JCAP {\bf 07} (2008) 013 [arXiv:0806.2911]
 \bibitem{F_sym} V. Vitale and A. Morselli for the Fermi/LAT Collaboration,
2009 Fermi Symposium [arXiv:0912.3828]

 \bibitem{GC_cim} 
 A. Morselli, B.Ca\~nadas,   V.Vitale,   Il Nuovo Cimento {\bf 34} C, N. 3 (2011)   [arXiv:1012.2292]
	
\bibitem{Halo}  M.Ackermann  {\it et al.} [Fermi Coll.],  ApJ  {\bf 761} (2012) 91   [arXiv:1205.6474]
 \bibitem{Fermi_el_interpretation}
D.~{Grasso}, S.~{Profumo}, A.~W. {Strong}, L.~{Baldini}, R.~{Bellazzini}, E.~D.
  {Bloom}, J.~{Bregeon}, G.~{di Bernardo}, D.~{Gaggero}, N.~{Giglietto},
  T.~{Kamae}, L.~{Latronico}, F.~{Longo}, M.~N. {Mazziotta}, A.~A. {Moiseev},
  A.~{Morselli}, J.~F. {Ormes}, M.~{Pesce-Rollins}, M.~{Pohl}, M.~{Razzano},
  C.~{Sgro}, G.~{Spandre} and T.~E. {Stephens}, {\em Astroparticle Physics} 
  {\bf 32} (2009) 140   [arXiv:0905.0636]

\bibitem{pamela_el_2011} O.Adriani. {\it et al.}~[PAMELA Coll.], Phys. Rev. Lett.  {\bf 106} ( 2011)  201101 
\bibitem{Fermi_el}   A.A.Abdo {\it et al.} [Fermi Coll.], PRL {\bf 102} (2009) 181101   [arXiv:0905.0025]
\bibitem{Fermi_el2}    M.Ackermann {\it et al.} [Fermi Coll.], Phys. Rev. D {\bf 82} (2010) 092004   [arXiv:1008.3999] 

\bibitem{Dwarf} A.Abdo {\it et al.} [Fermi Coll.],  ApJ  {\bf 712} (2010) 147-158     [arXiv:1001.4531]


\bibitem{Dwarf2}  M.Ackermann  {\it et al.} [Fermi Coll.], Phys. Rev. Lett.  {\bf 107}  (2011) 241302 [arXiv:1108.3546]

\bibitem{nfw} J.Navarro, J.Frenk,  S.White  {Astrophys. J.} {\bf 462}  (1996) 563  [arXiv:astro-ph/9508025]

	
 \bibitem{ani2} M.Ackermann  {\it et al.} [Fermi Coll.],  Phys. Rev. D {\bf 85} (2012) 083007 [arXiv:1202.2856]


\bibitem{DarSat} M.Ackermann  {\it et al.} [Fermi Coll.],  ApJ  {\bf 747} (2012) 121    [arXiv:1201.2691]



\bibitem{DarkSun}  M.Ajello  {\it et al.} [Fermi Coll.],      Phys. Rev. D {\bf 84}  (2011) 032007  [arXiv:1107.4272]

\bibitem{lines2} M.Ackermann {\it et al.} [Fermi Coll.],   Physical Review D {\bf 86} (2012) 022002     [arXiv:1205.2739] 


\bibitem{lines} A.Abdo {\it et al.} [Fermi Coll.], Phys. Rev. Lett. {\bf 104}  (2010) 091302  [arXiv:1001.4836]



\bibitem{lineW}   C. Weniger, JCAP {\bf 1208} (2012) 007 [arXiv:1204.2797 [hep-ph]].
\bibitem{lineF}    M. Su and D. P. Finkbeiner, arXiv:1206.1616 [astro-ph.HE].



\bibitem{Einasto}   T. Bringmann and C. Weniger, Dark Universe 1 (2012) 194-217  [arXiv:1208.5481]


\bibitem{line} 
 M.Ackermann {\it et al.}  [Fermi Coll.],   PRD     Submitted   [arXiv:1305.5597].
 



 

\bibitem{HEAT2004}  M. A. DuVernois {\it et al.},  [HEAT Coll.], ApJ {\bf 559} (2001) 296 
\bibitem{AMS2007}  M. Aguilar {\it et al.}, [AMS Coll.], Physics Reports {\bf 366} (2002) 331 


\bibitem{anis}   M.Ackermann {\it et al.} [Fermi Coll.], Phys. Rev. D  {\bf 82}, 092003 (2010)   [arXiv:1008.5119]

\bibitem{Fermi_elpos} M.~Ackermann {\it et al.}  [Fermi Coll.], Phys.\ Rev.\ Lett.\  {\bf 108}, 011103 (2012) [arXiv:1109.0521 [astro-ph.HE]].


\bibitem{AMS-02}  M. Aguilar et al. [AMS-02 Coll.]  PRL 110, 141102 (2013) 




\end{thebibliography}
\end{document}